# Measuring the interaction force between a high temperature superconductor and a permanent magnet

S. O. Valenzuela, G. A. Jorge, and E. Rodríguez

*Laboratorio de Bajas Temperaturas*

*Departamento de Física, Facultad de Ciencias Exactas y Naturales*

*Universidad Nacional de Buenos Aires*

*Ciudad Universitaria, Pabellón I*

*1428 Buenos Aires, República Argentina*

**Abstract**

Repulsive and attractive forces are both possible between a superconducting sample and a permanent magnet, and they can give place to magnetic levitation or free-suspension phenomena, respectively. We show experiments to quantify this magnetic interaction which represents a promising field regarding to short-term technological applications of high temperature superconductors. The measuring technique employs an electronic balance and a rare-earth magnet that induces a magnetic moment in a melt-textured $YBa_2Cu_3O_7$ superconductor immersed in liquid nitrogen. The simple design of the experiments allows a fast and easy implementation in the advanced physics laboratory with a minimum cost. Actual levitation and suspension demonstrations can be done simultaneously as a help to interpret magnetic force measurements.

# I. INTRODUCTION

The surge of superconducting compounds with critical temperatures, $T_c$, lying above liquid nitrogen boiling temperature[1] (77 K) has opened a new chapter in the field of solid state physics. The challenge to understand the properties of these "high temperature" superconductors (HTS) has inspired an amazingly active theoretical and experimental line of research since their discovery 12 years ago[2-4]. Current availability of these compounds tailored in the form of bulk ceramics, single crystals or thin films, has made possible the strengthening of the experimental treatment of superconductivity phenomenon in the undergraduate and advanced physics laboratory because liquid nitrogen is relatively inexpensive and much easier to handle than liquid helium and it requires far simpler cryogenic designs.

Experiments with HTS in the advanced physics laboratory have been proposed to detect a superconducting transition upon cooling a sample from the normal state[5-10]. In the more standard experiments, the electrical resistance of a sample[5-7], $R$, or its magnetic susceptibility[10], $\chi$, are measured to identify the key premises of superconducting behavior: zero electrical resistance and (perfect) diamagnetic response, respectively[11,12].

Beyond this first approach to the phenomenology of HTS, we believe a second approach should be presented through experiments evidencing the relevance of potential applications in connection with irreversible properties of these new materials. We present here experiments to analyze, in a quantitative way, the levitation phenomena and the free-suspension counterpart both resulting from the interaction of a superconducting sample with a permanent magnet. The analysis is done by measuring the interaction force between a HTS and the magnet by means of an electronic balance when the HTS is cycled in the magnet's field. In spite of the requirement of cryogenic liquid for cooling the samples below $T_c$, it is not necessary to control temperature and then using a cryostat is not imperative. Since experiments use conventional elements easily found in educational laboratories, they are implemented with no difficulty and with a minimum cost. These experiments may be complemented with demonstrations like those proposed by Early *et al.*[13] to help students to interpret the magnetic force measurements with actual observation of the magnetic levitation and suspension phenomena.

After introducing the main properties of superconducting materials and the concept of magnetic levitation and suspension associated with them, we will describe the experimental array and the measurements will be shown and discussed.

# II. MAIN CONCEPTS

Levitation or free-suspension of a body is possible if a force acts against gravity compensating the body's weight. Levitation may be attained by different methods (by a jet of air, by acoustic pressure, etc.)[14], although free suspension is somewhat more exotic. Stability is the main problem in the two cases. This condition, however, is fulfilled in the case of type-II superconductors interacting with a magnet, allowing the observation of both phenomena[15]. In such case, levitation or free-suspension of a superconducting body occurs with respect to the source of a non-uniform magnetic field.

These phenomena are both of academic and technological concern. From the point of view of possible applications, levitation of a superconductor above a magnet (or vice versa) is of central interest with regard to the commercialization of HTS. Indeed, magnetic levitation involving HTS is considered as a way to support high-speed vehicles[14,16] and some proposals and prototypes of trains levitated by superconducting coils actually exist[14,17]. It also appears possible to use these materials for magnetic bearing applications, such as generators, energy storage systems, and electric motors[18].

Having these applications in mind, superconductors should be capable to levitate with different objects attached to them and then, the interaction force with the magnet should be (much) higher than the superconductor's weight.

In this section we will survey the main concepts on the magnetic properties of superconductors that are necessary to understand the context for magnetic levitation involving superconducting materials.

### A. Meissner effect and vortices in superconductors

Any bulk superconductor immersed in a small enough magnetic field will create non-dissipative electrical currents confined near its surface that generates a magnetic field that opposes and cancels the externally applied field in the bulk material[19]. Thus a bulk superconducting material can exclude the external magnetic field to produce $B = 0$ in its interior (perfect diamagnetism) if the field is kept in a specific range of strength. This behavior is known as Meissner effect[19].

Type-I superconductors present a Meissner effect up to a thermodynamic critical field $H_c$ at which superconductivity is destroyed and the field fully penetrates into the bulk. Above this field the material is in the normal state.

On the contrary, in type-II superconductors the Meissner effect is only observed for magnetic fields below a lower critical field $H_{c1}$ and superconductivity is destroyed for fields above an upper critical field $H_{c2}$ ($H_{c1} < H_{c2}$). For magnetic fields between this two critical fields, the material is not perfectly diamagnetic. Indeed, the field penetrates into the bulk in association with supercurrent vortices that surround a normal core and carry a quantum of flux $\phi_0 = h/2e = 2.07\ 10^{-15}$ Wb. Each of these entities is usually termed flux line or vortex and when they are present the superconductor is said to be in the mixed or Shubnikov phase[20].

Most pure elements are type-I superconductors while most alloys and all HTS belong to the type-II family[21].

As we shall see stable levitation and suspension, in our case, are mainly due to hysteretical forces resulting from the interaction of flux lines with defects in the material.

### B. Pinning centers and critical current.

When a current flows in the presence of flux lines in an ideal type-II superconductor the flux lines start moving under the action of a Lorentz-type force $F_L$. However, this is not usually what happens in real type-II materials because static disorder due to lattice imperfections favors vortex pinning. The reason for this to occur is that, at places where a defect exists, the superconducting order parameter is naturally depressed and thus it is energetically favorable for a vortex to be in there in order to avoid wasting condensation energy for the formation of its normal core. Effective defects for vortex pinning in HTS may be of many types[22,23]: impurities, vacancies, dislocations, twins, stacking faults, local defects, added non-superconducting phases to the superconducting matrix, columnar defects created by irradiation with high-energy ions, etc.

In the presence of one or more of these types of defects, $F_L$ is counteracted by a pinning force, $F_p$. The Lorentz force per unit volume is given by $F_L = j \times B$, where $B$ is the magnetic induction (locally averaged over the flux lines) and $j$ is the current density. The maximum dissipation-free current is given by $j_c = F_p / B$. The critical current density $j_c$ is the de-pinning current density and it is a measure of the strength of pinning by static disorder in the material.

When an external field is applied, vortices do not distribute homogeneously in the sample because, as we have just pointed out, a critical force $j_cB$ is required to move the flux lines. This out-of-equilibrium distribution of vortices is equivalent to a magnetic field gradient which, by Ampère's law ($\nabla \times B = \mu_0 j$), leads to a bulk current density $j$.

A significant consequence of the presence of pinning centers is that the magnetization depends on the magnetic history of the superconducting sample, presenting hysteretic behavior and associated irreversibility when the sample is cycled in a magnetic field.

In the sixties C. P. Bean[24] computed the mean magnetization on a superconducting cylinder assuming a constant critical current density $j_c$ and that any modification in the flux distribution is introduced at the sample surface. Reversible components of the magnetization are not taken into account and $H_{c1}$ is supposed to be zero[24,25].

Again by Ampère's law, the field profiles in Bean's model are simply straight lines of slope $\partial B/\partial x = \mu_0 j_c$. We will come back to this model when we analyze the force measurements in section IV.

### C. Thermal activated movement of flux lines. Flux creep.

The concept of thermal activation of flux lines, together with the concept of pinning itself, were first introduced by P. W. Anderson in 1962 (Ref. [26]). The basic idea is that any process that promotes the out-of-equilibrium configuration of vortices to relax will lead to a change in the magnetic moment of the superconductor[26,27]. Such change can be thought to be a consequence of spontaneous creeping of vortices out of the pinning centers. This motion usually originates from thermal activation[26] (the vortex jumps over the pinning barrier), but it can also arise out of quantum tunneling[28] (at very low temperatures $T < 4$ K) or mechanical activation[29] as a consequence of vibrations (which in an applied field induce weak alternating currents).

This creeping motion is driven by a magnetic field gradient which leads to driving currents. As the configuration of vortices relaxes, the associated currents become smaller and the relaxation slows down giving place to an approximately logarithmic relaxation as first observed by Kim *et al.*[30].

### D. Hysteretical forces.

In general, the force $F$ over a distribution of currents $j$ due to an external magnetic field $H$ will be given by[31]:

$$F = \mu_0 \int j \times H \ dV \quad (1)$$

If the currents are in a volume $V$ where the field is sufficiently uniform so that we are allowed to assign a homogeneous magnetization $M$ to the currents and, furthermore, these currents loop in a plane perpendicular to the field then, from Eq. (1), the force will be parallel to the external field, let's say in $z$ direction, and will be given by[32]:

$$F_z = \mu_0 M_z \frac{\partial H_z}{\partial z} V \quad (2)$$

Therefore, to obtain a force due to the interaction of a superconductor with a magnetic field source, the superconductor should be located in a region of non-zero field gradient. As pinning leads to hysteretical magnetization, the force on the superconductor will also display hysteretical features.

An appropriate magnetic field to investigate this interaction force may easily be obtained with a commercial rare-earth permanent magnet. Inexpensive small-volume Sm-Co or Nd-Fe-B magnets are commercially available and provide fields in the range of 0.3-0.7 T close to their surface. For a given permanent magnet $\partial H_z/\partial z$ in Eq. (2) is fixed, and the sign of $F_z$ will depend on the sign of $M_z(H)$.

For high-temperature superconducting samples with strong enough pinning, levitation above or suspension below a magnet is possible; levitation being a consequence of flux exclusion ($M_z < 0$) and suspension of flux trapping inside the sample ($M_z > 0$).

## III. FORCE MEASUREMENT

Measurement of the interaction force between a HTS and a magnet are performed with the apparatus described in Fig. 1. We use an electrobalance (Sartorius BP 210 S) with 0.1 mg resolution and 210 g capacity (since electronic balances use strain gauges they actually measure force). On the measuring pan of the balance we attach upside-down a 10 cm-height styrofoam coffee cup where an iron washer is glued with varnish. A permanent magnet is magnetically fixed to the ring. The reason for using this arrangement is that the magnet has to be distant from the sensitive components of the electrobalance. Components of the balance are supposed to be not magnetic, although in our case, for instance, the metallic pan shows a slight magnetic behavior as deduced from a weak attraction towards the magnet observed when they are in close proximity.

We use a cylindrical Nd-Fe-B magnet (radius = 12 mm, height = 10 mm) with a field $\mu_0 H = 0.5$ T at the plane surface that decays down to about $10^{-3}$ T on the pan. Sensitive parts of the instrument are located below the measuring pan. Care must be taken to center the pair magnet-ring with respect to a vertical axis pointing to the middle of the circular measuring pan to accomplish the optimum operation conditions of the instrument.

A superconductor pellet is properly fixed inward a second small plastic coffee cup at the bottom's center. A complete filling of this glass with liquid nitrogen ($\approx 20$ cm$^3$) keeps the sample immersed at $T_{N2} = 77$ K $< T_c$ for approximately 15 minutes; enough time to carry out the necessary measurements. The glass is rigidly attached to a non-magnetic arm that can be moved vertically by means of a transmission gear to change the distance between the sample and the magnet. A ruler graduated in millimeters is used for measuring the distance between them. The levitation force is measured moving the sample at constant speed relative to the magnet, typically at 0.2 mm/s.

The set magnet-ring-glass loads the balance with ~70 g, and this weight is tared to have a null starting reading. In these conditions, the balance will sense as an extra load the force due to the magnetic interaction between the superconductor and the magnet, and the instrument will give positive or negative readings as a signature of the repulsive or attractive character of this interaction. The null reading may correspond to the interaction force when *i*) the bulk sample is in the superconducting state but "infinitely" away from the magnet, *ii*) the total magnetic moment is averaged to zero in the volume of the sample, *iii*) the material is at a $T > T_c$ and displays normal properties.

Since the experiments can be completed in a few minutes, we prefer not to perform data acquisition via the available RS232 interface of the balance and thus the interaction force is obtained by directly reading its display. If wanted a complete automatization of the acquisition process is possible using the RS232 interface and a step motor to control the movement of the sample.

The sample used is a melt-textured $YBa_2Cu_3O_7$ ($T_c$=92K) with cylindrical shape (radius = 3 mm, height = 2.5 mm, weight $w$ = 4.1 $10^{-3}$ N ). The sample contains micrometric inclusions of non-superconducting $Y_2BaCuO_5$ incorporated to the superconducting matrix during the growing process in order to increase the number of pinning centers[33]. This results in an enhanced critical current density in relation to pure superconducting $YBa_2Cu_3O_7$ ceramics. A typical value of $j_c \approx 10^9$ A/m$^2$ at 77 K and zero magnetic field is measured by magnetization experiments[34]. Irreversible behavior of this compound is observed below an "irreversibility line" described by $\mu_0H_{irr} = \mu_0H_0$ (1 - $T$ / $T_c$)$^n$, with $\mu_0H_0 \approx 60$ T and $n \approx 3/2$ (Ref. [34]). This assures that our sample will display hysteretical features related to its magnetic history at liquid nitrogen temperature in presence of the maximum field provided by our permanent magnet, $\mu_0H_{max} = 0.5$ T $<<$ $\mu_0H_{irr}$(77K) $\approx 4$ T.

## IV. RESULTS

Fig. 2 shows typical results obtained with the setup described in section III. Fig. 2a) shows the axial force when the sample is cooled from above $T_c$ in zero magnetic field (zero field cooled, ZFC) and then cycled in the field of the magnet as indicated by the arrows. Fig. 2b) shows the axial force when the sample is cooled from above $T_c$ in close proximity to the magnet (field cooled, FC). Here $z$ is measured from the upper face of the magnet.

The interaction force $F(z)$ depends on the product of the magnetization $M(z)$ and the field gradient $H'(z)$ so that, at each relative position, we have $F(z) \propto M(z) H'(z)$ as indicated by Eq. (2) (note that $F(z)$ implicitly depends on $H$ through $z$). If both the magnetic field and the magnetic field gradient are known, we can invert Eq. (2) to obtain $M(H)$ from the measured force.

The magnetic field supplied by the magnet was measured with a Hall probe[35] (inset in Fig. 2a)). The field gradient can be obtained both numerically from the magnetic field dependence or experimentally using the same setup in Fig 1 and measuring the interaction force between the magnet and a known magnetic dipole $m$ ($H'(z) \propto F(z)/m$).

We measured the field gradient using a 5 mm diameter coil of 20 turns wound with 100 µm copper wire with an applied current $I$ = 100 mA ($m$ = $I N A$ = 39.3 $10^{-6}$ A m$^2$, where $A$ is the area of the coils and $N$ the number of turns). See the inset in Fig. 2b).

The magnetization values $M(H)$ related to the ZFC measurements in Fig. 2a) are presented in Fig. 3. We will begin analyzing this curve to gain an insight into the magnetization behavior when the sample is cycled in the field of the magnet. After that, we will return to the interpretation of the measured interaction force in Fig 2.

### A. Hysteretic magnetization loop

The specific relationship between $M$ and $j_c$ depends on the sample geometry (see, for example, Refs. [36-38]). Here, for illustrative purposes, we will use the results of a simple case to qualitative explain the shape of the hysteresis loop: a slab of thickness $2d$ in an applied field parallel to its surface (*i.e.*, no demagnetizing factors are considered) in the framework of the Bean model.

We will analyze the ZFC case but an analogous analysis can be done for the FC data. The arrows in Fig. 3 indicate how the sample was cycled.

Initially, the sample is far away from the magnet and then it is drawn near it, therefore the initial field is zero and then starts to increase. As the field increases, flux penetrates into the sample from its surface with a gradient $\partial B/\partial x = \mu_0 j_c$ to a distance $x = H / j_c$. The field profile is sketched at point 1 in Fig.3. The magnetic moment per unit sample volume, given by $M = <B>/\mu_0 - H$, is:

$$M = \frac{H^2}{2 j_c d} - H \qquad (3)$$

reaching its maximum (diamagnetic) value when the flux fronts penetrate to the center ($H = H^* = j_c d$):

$$M = -\frac{j_c d}{2} \qquad (4)$$

This occurs around point 2 as represented in Fig. 3.

For a disk-like geometry (in an axial field) the maximum value of $M$ is[38]:

$$M = -\frac{1}{3} j_c R \qquad (5)$$

with $R$ being the sample radio.

As the field is reduced back to zero (the sample is moved away from the magnet), the field gradient changes sign at the surface (*e. g.*, at points 3 and 4) and the magnetic moment becomes positive (paramagnetic) giving rise to a remanent magnetization. This remanent magnetization originates from flux trapping inside the superconductor. We have $M = 0$ when the total magnetic moment is averaged to zero in the volume of the sample (point 3).

From Eq. (5) (and also from $H^*$) we can estimate the critical current which is in agreement with previously reported values in similar applied field and temperature [34]: $j_c \approx 9\ 10^7$ A/m$^2$. Nevertheless, this value can be considered only as a rough approximation. There are three main reasons for that, *i*) we have not taken into account demagnetizing factors, *ii*) the field is not uniform over the sample (see the inset in Fig. 2 a) and *iii*) thermally activated movement of flux lines changes the magnetization with time.

Magnetic relaxation is indeed observed beyond point 5 in Fig. 3. In experiments in which the magnetic field is cycled, like in this case, the creeping motion of vortices is usually in competition (at the sample surface) with the rate of change of the magnetic field. As the sample is moved away from the magnet at constant speed, the rate of change of the magnetic field is greatly reduced and the relaxation should become noticeable. The Bean model predicts the magnetization should increase (or at least be constant) as the field is reduced but, what is actually found, beyond point 5, is the opposite behavior. In fact, magnetic relaxation can be readily observed and studied with the setup of the present experiment by fixing the relative position between the magnet and the superconductor and observing the force as a function of time. This will be the central topic of a forthcoming article.

**B. Hysteretical forces**

We have discussed how flux pinning produces an irreversible magnetization, thus we are able to analyze the hysteretical force *F* in Fig. 2. It is worth noting that the following analysis can be done accompanied with simultaneous demonstrations of the studied phenomena.

For levitation or suspension to be possible, *F* must compensate the superconductor's weight *w*, *i.e.* $F = w$ must be satisfied. The weight of the sample is $w = 4.1 \times 10^{-3}$ N, a value that is much smaller than the maximum measured interaction force and can hardly be plotted on the scale of Fig 2. Note that the maximum force displayed is almost one hundred times the weight of the sample. For clarity, we will consider the sample having a load attached to it and will discuss when levitation or suspension is possible in this case. The load is supposed to be $w_L = 15\,w$.

As shown in Fig. 3, when the sample is zero field cooled (ZFC) it initially has $B = 0$ and $M = 0$ and then $F = 0$. If it comes nearer the magnet, it gains negative magnetization by preventing the vortices from penetrating freely inside (the vortices got pinned near the surface). The sample feels a repulsive force and the pair sample-load reaches stable levitation at point A in Fig. 2a) (*i.e.*, $F = w_T = w + w_L$ and $dF/dz < 0$ are satisfied).

When the pair is pushed even closer to the magnet, both the absolute magnetization of the sample and the field gradient grow. At some point between A and B, the magnetization reaches its maximum value and further on, the measured force is proportional to the field gradient. So the force keeps growing and greatly exceeds $w_T$ at, for instance, point B. If the loaded sample where released at this point, then stable levitation would be obtained at point C. By closely examining $F(z)$ and the magnetization curve, one can easily be convinced stable levitation is possible at any height between C and A by releasing the loaded sample at a point farther from the magnet than point B.

If now we imagine the loaded sample to be below the magnet then, after being at point B, it would reach stable suspension at point D (magnetization changes sign when $F = 0$). Again, examination of the behavior of $F(z)$ shows stable suspension is possible between D and E (note that E is not stable because $dF/dz > 0$).

A similar interpretation of the force behavior in the FC case may be given. In this case, the sample is cooled 2 mm above the magnet from $T > T_c$ to $T \approx 77$ K $< T_c$ (point A' in Fig.2 b)).

After being cooled in the field the superconductor behaves as a diamagnetic with a small negative magnetization. This results in a non-zero force measured by the balance. This magnetization is reversible and is not related with pinning[15]. If the loaded sample were free to move it would achieve stable levitation or suspension at points B' or C', respectively. Here irreversible properties dominate again. Note that if the sample were cooled resting at the magnet surface it might also lift the load up to a stable position due to the mentioned reversible magnetization.

Finally, stable suspension is possible between points C' and D' (D' is not stable). When the sample is moved away from the magnet and then approached to it, stable levitation is achieved at point E'. The force is not strong enough for suspension as the maximum measured attractive-force is about 0.045 N $< w_T$.

## V. FINAL REMARKS

In summary, we have presented an experiment to quantify the interaction between a HTS sample and a magnet. The experiment employs an electronic balance, a rare-earth magnet, a melt-textured $YBa_2Cu_3O_7$ superconductor and liquid nitrogen. After measuring the force as a function of the relative position, $F(z)$, with the balance, we estimated the magnetization of the superconductor, $M(H)$. The analysis of $M(H)$ in terms of pinning of vortices in a type-II superconductor allowed us to give a qualitative explanation of the features of the force curve.

We would finally like to note that a clear observation of the change in the sign of $F(z)$, when cycling the sample in the field of the magnet, is possible for a material like a melt-textured $YBa_2Cu_3O_7$ with normal inclusions, having a high critical current density $j_c$. If the experiments were run using a pure polycrystalline $YBa_2Cu_3O_7$ sample where weak links between grains dominate the magnetic behavior, the force curve might not display strong hysteretical features as in Fig. 2. This will be a direct consequence of a deeper field penetration in the bulk due to a lower $j_c$. In this case a narrow $F(z)$ loop would be expected or, in the case of an extremely low $j_c$, a nearly reversible $F(z)$ in all ZFC or FC processes. For the same reason, if the force experiments using a ceramic were accompanied with *in situ* demonstration of the levitation and suspension phenomena, it might be difficult to observe levitation at 77 K (depending on the strength of the magnet), while suspension might be even impossible to attain.

## ACKNOWLEDGEMENTS

The authors are grateful to X. Obradors and S. Piñol (Institut de Ciència de Materials de Barcelona, CSIC) for providing the melt-textured $YBa_2Cu_3O_7$ sample and colleagues of Laboratorio de Bajas Temperaturas for a critical reading of the manuscript. S.O.V. thanks University of Buenos Aires (UBA) and Fundación Sauberán, and E.R. thanks FOMEC-UBA for financial support. G. A. J. is an undergraduate Physics student.

**REFERENCES**

[1] M. K. Wu, J. R. Ashburn, C. J. Torng, P. H. Hor, R. L. Meng, L. Gao, Z. J. Huang, Y. Q. Wang, and C. W. Chu, "Superconductivity at 93 K in a new mixed-phase Y-Ba-Cu-O compound system at ambient pressure," Phys. Rev. Lett. **58**, 908-910 (1988).

[2] T. G. Bednorz and K. A Müller, "Possible high $T_c$ superconductivity in the Ba-La-Cu-O system," Z. Phys. B **64**, 189-193 (1986).

[3] Charles P. Poole Jr., Horacio A. Farach and Richard J. Creswick, *Superconductivity* (Academic Press, San Diego, 1995).

[4] G. Burns, *High-Temperature Superconductivity, an Introduction* (Academic Press, San Diego, 1992), pp. 1-7.

[5] L. M. León-Rossano, "An inexpensive and easy experiment to measure the electrical resistance of high-$T_c$ superconductors as a function of temperature," Am. J. Phys. **65**(10), 1024-1026 (1997).

[6] K. G. Vandervoort, J. L. Willingham and C. H. Morris, " Simple, inexpensive probe for resistivity measurements above 77 K on metals and superconductors," Am. J. Phys. **63**(8), 759-760 (1995).

[7] Michael J. Pechan and Jonathan A. Horvath, "Quasiequilibrium determination of high-$T_c$ superconductor transition temperatures," Am. J. Phys., **58**(7) 642 - 644 (1990).

[8] H.G. Lukefahr, V. Priest, K.B. St. Jean, J.S.R. Worley, and C.S. Yeager D.A. Gajewski and M.B. Maple, "A very simple and inexpensive apparatus for detecting superconducting transitions via magnetic screening," Am. J. Phys. **65**(2), 132-135 (1997).

[9] J. N. Fox, F. A. Rustad and R. W. Smith, "Measurement of the transition temperature of a high-$T_c$ superconductor," Am. J. Phys. **56**(11) 980-982 (1988).

[10] Martin Nikolo, "Superconductivity: A guide to alternating current susceptibility measurements and alternating current susceptometer design," Am. J. Phys. **63**(1), 57-65 (1995).

[11] M. Thinkam, *Introduction to Superconductivity* (Mc Graw-Hill, New York, 1996), 2$^{nd}$ ed., pp. 2-9.

[12] P. G. de Gennes, *Superconductivity of metals and alloys* (Addison-Wesley, 1989), pp. 3-12.

[13] E. A. Early, C. L. Seaman, K. N. Yang and M. B. Maple, "Demonstrating superconductivity at liquid nitrogen temperatures," Am. J. Phys. **56**(7), 617-620 (1988).

[14] E. H. Brandt, "Levitation in physics," Science **243**, 349-355 (1989).

[15] E. H. Brandt, "Rigid levitation of high-temperature superconductors by magnets," Am. J. Phys. **58**(1), 43-49 (1990).

[16] R. L. Byer, R. F. Begley and G. R. Stewart, " Superconducting, magnetically levitated merry-go-round," Am. J. Phys. **42**(2), 111-125 (1974).

[17] See for example, Railway Technical Research Institute (RTRI) home page at http://www.rtri.or.jp or Donald B. Perkins (Maglev Central) home page at http://web.syr.edu/~dbperkin/MaglevCentral.html.

[18] J. R. Hull, E. F. Hilton, T. M. Mulcahy, Z. J. Yang, and A. Lockwood, "Low friction in mixed-mu superconducting bearings," J. Appl. Phys. **78**(11), 6833-6838 (1995).

[19] See Ref. 11, pp. 19-22 and Ref. 12, pp. 4-7.

[20] See Ref. 11, pp. 11-13 and Ref. 12, pp. 12-13 and 48-55.

[21] See Ref. 3, pp. 265-290.

[22] K. Salama and D. F. Lee, "Progress in melt texturing of $YBa_2Cu_3O_x$ superconductor," Supercond. Sci. Technol. **7,** 177-193 (1994).

[23] L. Civale, "Vortex pinning and creep in high-temperature superconductors with columnar defects," Supercon. Sci. Technol. **10,** A11-A28 (1997).

[24] C. P. Bean, "Magnetization of hard superconductors," Phys. Rev. Lett. **8**, 250-253 (1962).

[25] C. P. Bean, "Magnetization of high-field superconductors," Rev. Mod. Phys. **36**, 31-39 (1964).

[26] P. W. Anderson, "Theory of flux creep in hard superconductors," Phys. Rev. Lett. **9**, 309-311 (1962).

[27] Y. Yeshurun, A. P. Malozemoff and A. Shaulov, "Magnetic relaxation in high-temperature superconductors," Rev. Mod. Phys. **68**(3), 911-949 (1996).

[28] G. Blatter, V. B. Geshkenbein, and V. M. Vinokur "Quantum collective creep," Phys. Rev. Lett. **66**, 3297-3300 (1991).

[29] E. H. Brandt, P. Esquinazi, H. Neckel, and G. Weiss, "Drastic increase of frequency and damping of a superconducting vibrating reed in a longitudinal magnetic field," Phys. Rev. Lett. **56**, 89-92 (1986). E. Rodríguez, J. Luzuriaga, C. D'Ovidio, and C. Esparza, "Softening of the flux-line structure in $La_{2-x}Sr_xCuO_4$ measured by a vibrating reed," Phys. Rev. B **42**, 10796-10800 (1990).

[30] Y. B. Kim, C. F. Hempstead and A. R. Strand "Critical persistent currents in hard superconductors," Phys. Rev. Lett. **9**, 306-309 (1962).

[31] J. D. Jackson, *Classical electrodynamics* (John Wiley & Sons Inc., New York, 1975), 2$^{nd}$ ed., pp. 173.

[32] Ref. 31, pp. 184-185.

[33] S. Piñol, F. Sandiumenge, B. Martínez, V. Gormis, J. Fontcuberta, X. Obradors, E. Snoeck and Ch. Roucau, "Enhanced critical currents by $CeO_2$ additions in directionally solidified $YBa_2Cu_3O_7$," Appl. Phys. Lett **65**(11), 1448-1450 (1994).

[34] B. Martínez, X. Obradors, A. Gou, V. Gomis, S. Piñol, J. Fontcuberta and H. Van Tol, "Critical currents and pinning mechanisms in directionally $YBa_2Cu_3O_7$-$Y_2BaCuO_5$ composites," Phys. Rev. B **53**(5), 2797-2810 (1996).

[35] We used a miniature Hall probe THS 118 (GaAs) with a sensitive area of (100x100) $\mu m^2$. The sensor is encapsulated in a plastic holder of dimensions (1.2x1.2x0.4) $mm^3$. With an input dc current $I = 1$ mA, $V_{Hall} = 0.18$ V/T.

[36] A. M. Campbell and J. E. Evetts, "Flux vortices and transport currents in type-II superconductors," Adv. Phys. **21**, 199-428 (1972) .

[37] E. H. Brandt, "The flux-lattice in superconductors," Rep. Prog. Phys. **58**, 1465-1594 (1995).

[38] J. Gilchrist, "Critical state model: comparison of transverse and elongated geometries," Physica C **219**, 67-70 (1994). Detailed calculations of the flux penetration and the magnetization curves in type-II superconducting disks of finite thickness can be found in E. H. Brandt "Superconductor disks and cylinders in an axial magnetic field. I. Flux penetration and magnetization curves, " Phys. Rev. B **58**, 6506-6522 (1998).

**FIGURE CAPTIONS**

**Fig. 1:** Experimental array for the force measurement. (A) electrobalance, (B) measuring pan, (C) plastic glass, (D) washer, (E) permanent magnet, (F) plastic glass containing the superconducting sample, (G) superconducting sample, (H) mobile non-magnetic arm, (I) rule in mm.

**Fig. 2:** a) Measured force between the superconducting sample and the magnet in a zero field cooling experiment (ZFC). The dashed lines indicate the weight of a load, $w_L$ (attached to the superconductor) plus the weight of the superconductor interacting with the magnet, $w$ ($w_T = w + w_L$). Inset: Axial magnetic field supplied by the Nd-Fe-B magnet. b) Measured force in the case of a field cooling experiment (FC). Inset: Gradient of the axial magnetic field of the magnet. Labels A, B, C,..., A', B', C',... indicate relevant distances of the sample to the magnet (see text).

**Fig. 3:** Magnetization curve $M(H)$ associated to the ZFC experiment of Fig. 2a) obtained through Eq. 2. The pictures display the field profile inside the superconducting sample at different values of the applied magnetic field, the shaded areas indicate regions of flux penetration. The numbers 1-5 identify relevant points on the hysteresis curve (see text). From the saturation value of the magnetization we estimate a critical current $j_c \approx 9\ 10^7$ A/m$^2$.

Fig. 1, S. O. Valenzuela *et al.*, Am. J. Phys.

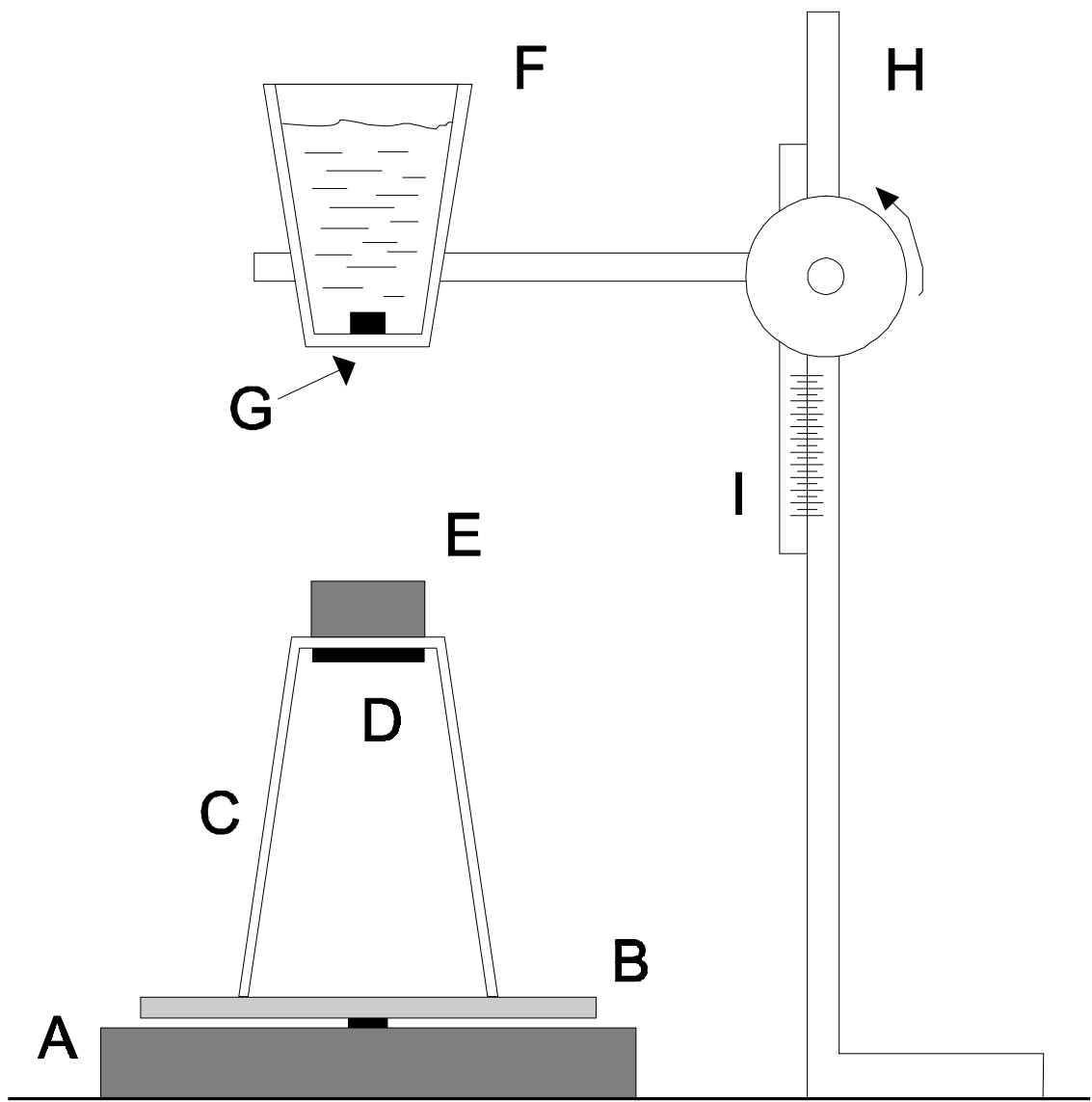

Fig. 2., S. O. Valenzuela *et al.*, Am. J. Phys.

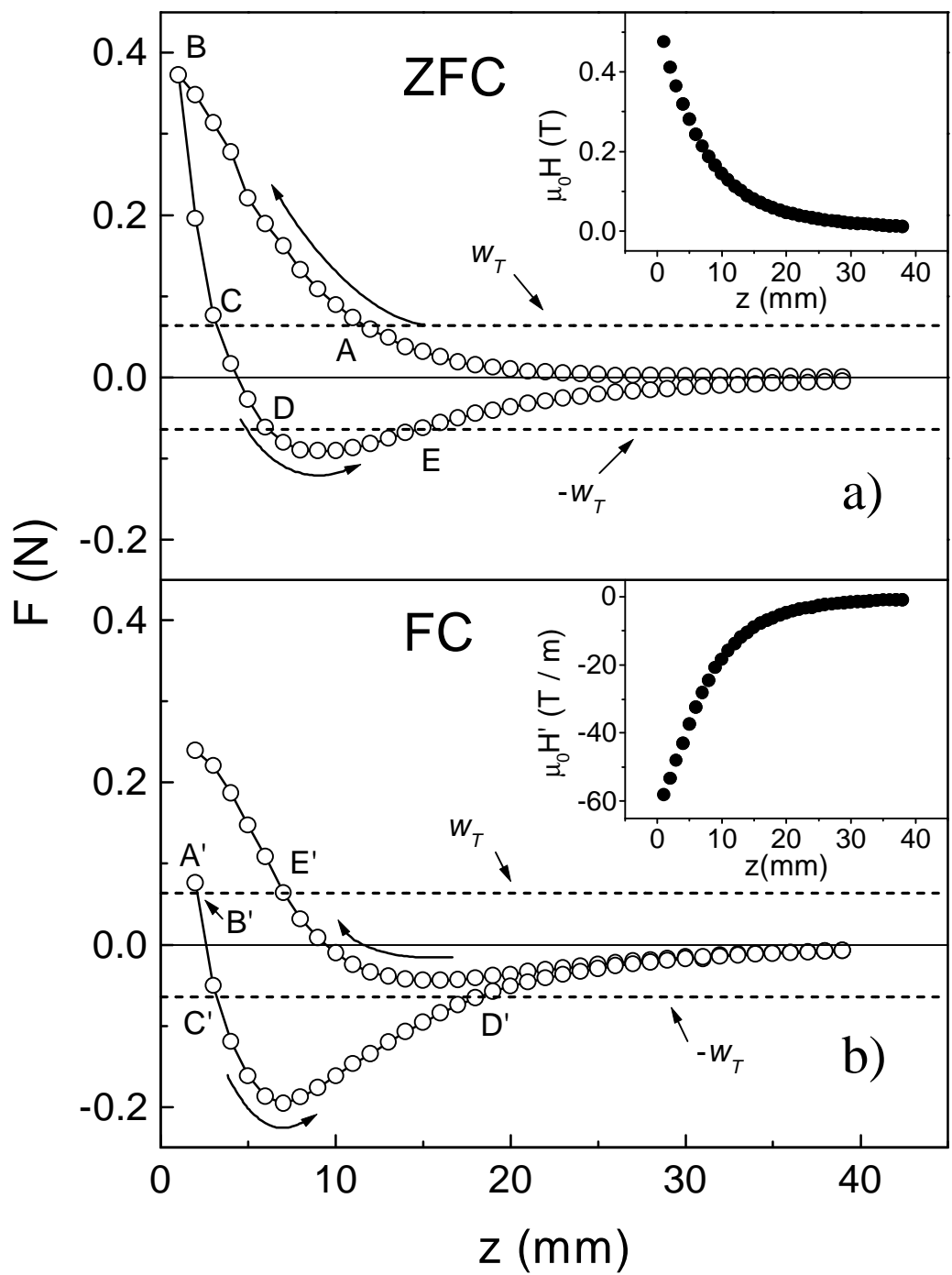

Fig. 3, S. O. Valenzuela *et al.*, Am. J. Phys.

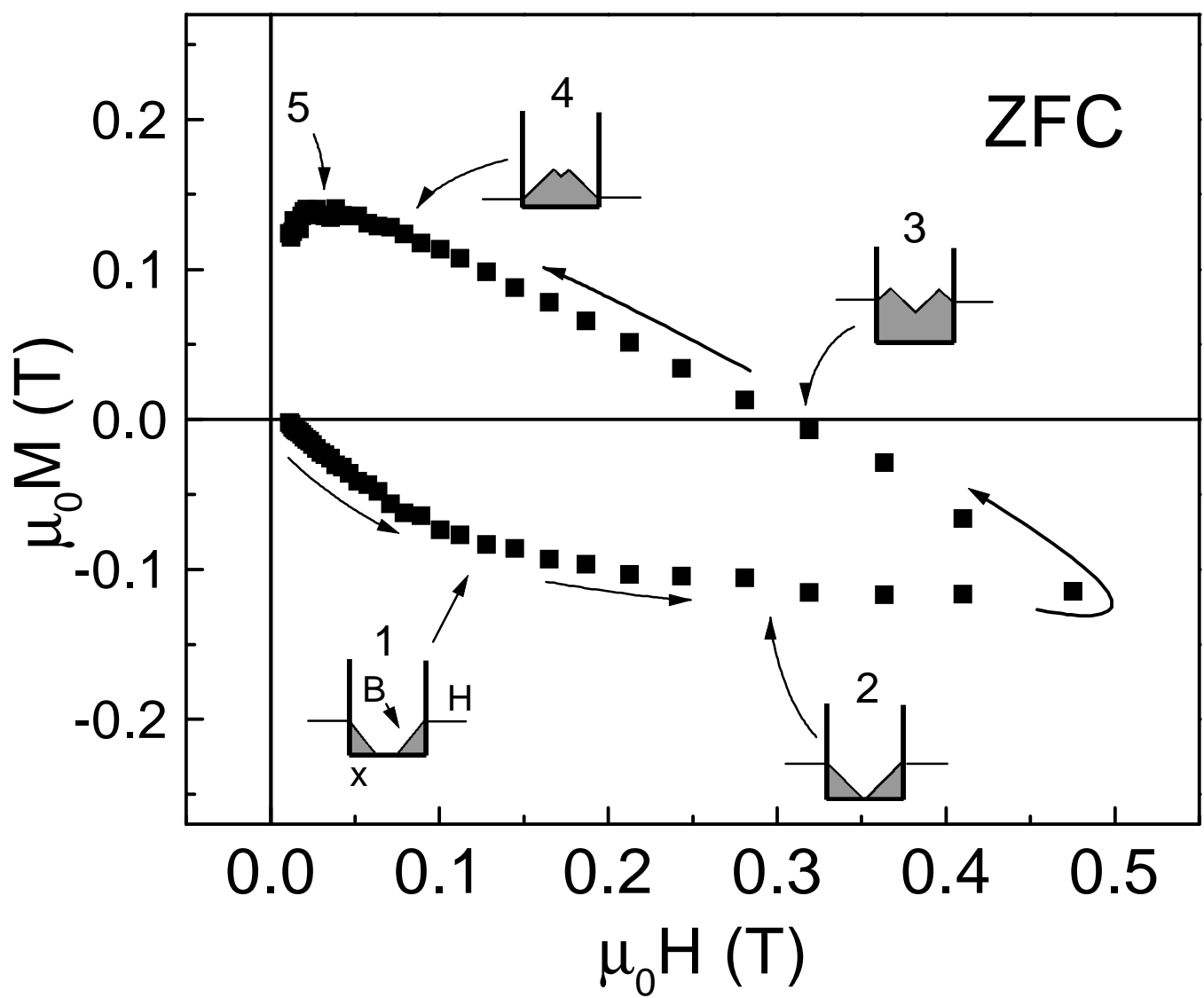